\pdfoutput=1
\documentclass{llncs}

\usepackage{url}
\usepackage{comment} 
\usepackage{graphicx}
\usepackage{diffs,xspace}
\RequirePackage{amssymb}
\usepackage{algorithm2e}
\RequirePackage[centertags]{amsmath}
\usepackage{multirow}

\newcommand{\mail}{\texttt{\{Ning.Ge | Marc.Pantel | Xavier.Cregut\}@enseeiht.fr}\xspace}

\newcommand{\vv}{\texttt{V\&V}\xspace}
\newcommand{\mde}{\texttt{MDE}\xspace}

\newcommand{\uml}{\texttt{UML}\xspace}
\newcommand{\marte}{\texttt{MARTE}\xspace}

\newcommand{\tpn}{\texttt{TPN}\xspace}

\newcommand{\bcet}{\texttt{BCET}\xspace}
\newcommand{\wcet}{\texttt{WCET}\xspace}

\newcommand{\dsl}{\texttt{DSL}\xspace}
\newcommand{\gpl}{\texttt{GPL}\xspace}
\newcommand{\simulink}{\texttt{Simulink}\xspace}
\newcommand{\scade}{\texttt{SCADE}\xspace}
\newcommand{\alf}{\texttt{ALF}\xspace}
\newcommand{\aadl}{\texttt{AADL}\xspace}
\newcommand{\eastadl}{\texttt{EAST-ADL}\xspace}

\newcommand{\dkl}{\texttt{Kullback-Leibler Divergence}\xspace}

\newcommand{\idf}{\texttt{IDF}\xspace}
\newcommand{\tf}{\texttt{TF}\xspace}

\newcommand{\tc}{\texttt{TC}\xspace}
\newcommand{\itc}{\texttt{ITC}\xspace}
\newcommand{\exam}{\texttt{EXAM}\xspace}

\newcommand{\hmm}{\texttt{HMM}\xspace}

\begin{document}
\title{Probabilistic Failure Analysis in Model Validation \& Verification}

\author{Ning Ge \and Marc Pantel \and Xavier Cr\'egut}

\institute{University of Toulouse, IRIT/INPT\\
\mail
}

\maketitle
\begin{keywords}
	Fault localization, Model checking, Verification, Validation, Probabilistic analysis, Hidden markov model
\end{keywords}

\section{Introduction}
\label{introduction}
\begin{figure}
	\centering
	\includegraphics[height=0.32\textheight]{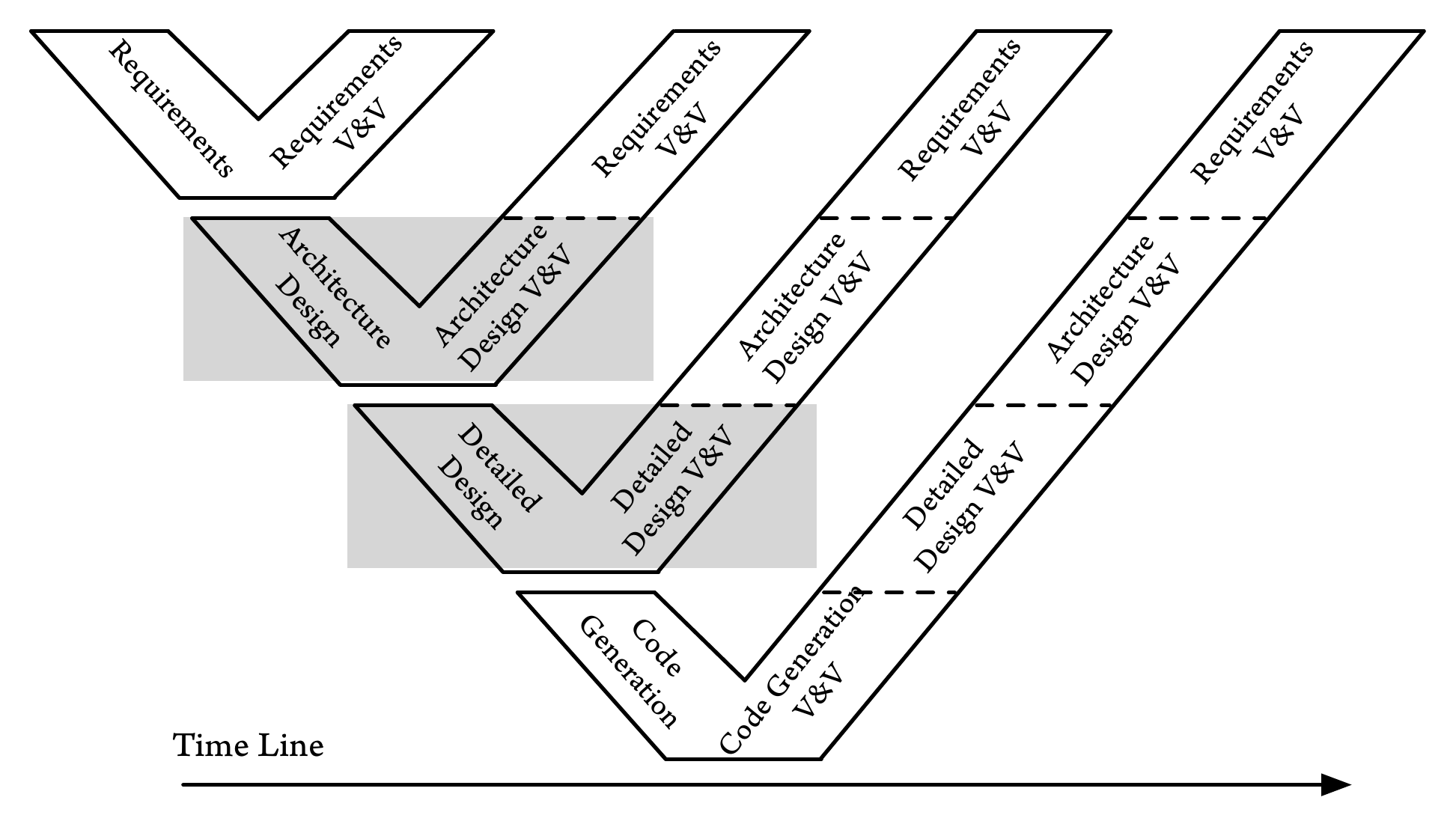}
    \caption{V-Model in Model-Driven Engineering}
    \label{fig:vmodel}
\end{figure}

As the size and complexity of safety critical real-time system are rapidly increasing due to the evolution of functional and non-functional requirements, Model-Driven Engineering (\mde) has become a promising means to improve the reliability and efficiency of the traditional software engineering by introducing the models and the formal methods. 
We use the multi V-model proposed in the \texttt{ITEA TIMMO} project in Fig. \ref{fig:vmodel} to illustrate the use of process of \mde for  developing real-time system. 
In order to generate reliable execution code, the verification and validation (\vv) are performed on each phase of system development lifecycle. The architecture design is the phase to design hardware and software architecture which can also be referred to as high-level design. It should involve a brief and abstract functionality of each module, their interface relationships, dependability, architecture diagrams, etc. The detailed design model can also be called modular or function design model, where the low-level design including detailed functional logic of the module can be specified. 
From the current practice, the architecture is usually modeled using Domain Specific Language (\dsl) such as \aadl and \eastadl or specific diagrams in a General Purpose Language (\gpl) such as \uml (Composite Structure Diagram),  while the detailed design is usually modeled using \dsl such as \simulink and \scade or specific diagrams such as \uml Activity, State Machine Diagrams, or \alf (Action Language for Foundational \uml). 

In practice, \vv in \mde is usually implemented in two manners: simulation and formal verification (such as  static analysis, theorem proving and model checking). 
When a requirement is not satisfied, the verification results will be used to diagnose this failed design. The end users will analyze the verification results to locate the origin of fault ( an activity called fault localization). The efficient and effective  failure analysis method in \vv is an important issue. 

Fault localization is not trivial in \mde. In the architecture design \vv, due to the use of abstraction for scalable model checking, some unnecessary information are removed for some specific \vv purpose. This reduces the state space explosion problem, but usually leads to model with concurrent and indeterministic behaviors that are  much more complicated to debug.
In the detailed design \vv, fault propagation is one of main issues. When a functional constraint is violated, we detect wrong output values from some components. Usually the others' outputs will be affected due to failure propagation. 

In order to locate the potential origin of faults from failure scenario in the model \vv, we propose to analyze fault locations using probabilistic approaches inspired by data mining technologies. Considering the different features of architecture design and detailed design, we use \dkl to analyze the failure from exhaustive model checking results in the architecture design \vv, and use hidden markov model to analyze the failure from  simulation in the detailed design \vv. This paper describes early experiments on these two aspects. 


\section{Probabilistic Failure Analysis in Model Checking}
\label{modelchecking}
\subsection{Problem Statements}
Generating a counterexample in case a formula is violated is a key service provided by model checkers. Counterexamples produced by model checkers often stand for error traces, which represent sequences of system states and transitions and are therefore usually lengthy and difficult to understand. The origin of error might be anywhere along these traces, thus requiring a lengthy analysis by designers. 
Our ultimate goal is to detect and to provide the end users with the suspicious ranked faulty elements. 
\subsubsection{Abstraction Issue}
Fault localization in model checking is challenging as the models have usually a concurrent and indeterministic behavior with many possible execution traces. This behavior is due to the use of abstraction in their design.  Without precise information, fault localization may not be precise enough. Given a sequential, or synchronized concurrent, program which exhibits less execution traces, various debugging methods are available to detect and locate the faulty statements. In model-based diagnosis, the use of abstraction is mandatory to reduce the state space explosion problem. At the time of writing, the conflict between model precision and verification cost is a key issue in model checking and model-driven engineering (\mde), therefore a compromise is made to remove the unnecessary information for some verification purpose while keeping all the property-related information. This usually leads to model with concurrent and indeterministic behaviors that exhibits a much larger number of execution traces and are consequently much more complicated to debug.

\subsubsection{Fault Localization Issue}
Sometimes it is difficult, even for seasoned experts, to analyze the fault origin. We take a simple example (see Ex. \ref{7ex:faultex}) to illustrate this issue.
\begin{example}[Fault Localization Example]
\label{7ex:faultex}
Assume a system consists of two concurrent processes $A$ and $B$. Both execute only once. The execution time is [5,10] for $A$, and [3,7] for $B$.  The expected temporal property $\mathcal{P}$ is \textbf{Always} $A$ \textbf{After} $B$. 

It is obvious that $\mathcal{P}$ is unsatisfied. The design fault occurs either on $A$ or on $B$. To remove this violation, we can either replace the time constraint of $A$ by [8,10], or replace the time constraint of $B$ by [3,4]. However, without extra information, $A$ and $B$ exhibit the same suspicion. If an extra information is available, e.g. the best case execution time (\bcet) of $B$ is 5, then the time constraint of $B$ cannot anymore be replaced by [3,4], thus the suspicion of $B$ is largely decreased.
\end{example}

This example is simple enough to be analyzed manually, while it is impossible for more complex system with thousands of transitions. Any modification on a transition may impact the verification result through time constraint propagation.

\subsubsection{Proposed Approach}

Existing automated fault localization techniques in model checking usually produce a set of suspicious statements without any particular ranking \cite{ball2003,groce2003,groce2004error,chaki2004explaining,jose2011cause}. 
Our approach will improve the effectiveness of fault localization by providing a suspiciousness factor which is used to rank the suspicious transitions in the verification model \cite{ge2014automated}. The suspiciousness factor is computed using the fault contribution of each transition on the error traces derived from the reachability graph. This approach has been applied in our formal verification framework of \uml-\marte designs dedicated to the real-time properties \cite{ge2012time,ge2012formal,ge2012transformation,ge2014reduction,ge2014uml}.
\subsection{Preliminaries}

\subsubsection{Reachability Graph \& Violation States}
Reachability graphs are used to solve reachability problem in model checking. They contain all the states in the execution of a system and all the transitions between these states. 
When a safety property is not satisfied, there exists violation states in the reachability graph. 
Finding all violation states in the reachability graph is the first step of error localization. 

\subsubsection{Fault Contribution of Transition}
\begin{definition}[Fault Contribution]
Fault Contribution ($C_F$) is a suspiciousness factor to evaluate a transition's suspicion level. It is used to rank the suspiciousness of transitions. 
\end{definition}

\subsubsection{Error Traces}
\begin{definition}[Error Trace]
\label{7def:errortrace}
For all the states \{$s_i$\} on the path from an initial state $s_0$ to a violation state $s_v$ in the reachability graph, all the outgoing transitions of $s_i$ are considered as error trace $\pi$.
\end{definition} 

We consider not only the transitions on the path that leads $S_0$ to $S_v$ in the definition of error trace but also the direct outgoing transitions of all the states in execution traces that lead to correct states. Indeed, in \tpn, the transitions outgoing from the same place can mutually influence each other. A faulty transition can change the way a correct transition is fired if they are both outgoings from the same place. The correct transition will diminish the $C_F$ of the faulty transition.

\begin{example}[Error Trace Example]
\label{7ex:errortrace}
In Fig. \ref{7fig:errortraceex}, $s_0$ is initial state, $s_v$ is a violation state. On the execution trace from $s_0$ to $s_v$, there exist four states \{$s_0, s_1, s_2, s_3$\} (apart from $s_v$). 
The state $s_3$ is in a correct trace. When the system is in state $s_2$, it is possible to transit to $s_7$ leading to a correct trace, or to $s_3$ leading to a violation state. If $s_7$ is removed from the graph, $s_3$ will have higher fault contribution for the violation state.  
The outgoing transitions of these four states are considered as error traces $\pi$, i.e., $\pi = \{t_0, t_1, t_2, t_1, t_5, t_4, t_2, t_3, t_4\}$.
\begin{figure}[ht!]
	\centering
		\includegraphics[height=0.11\textheight]{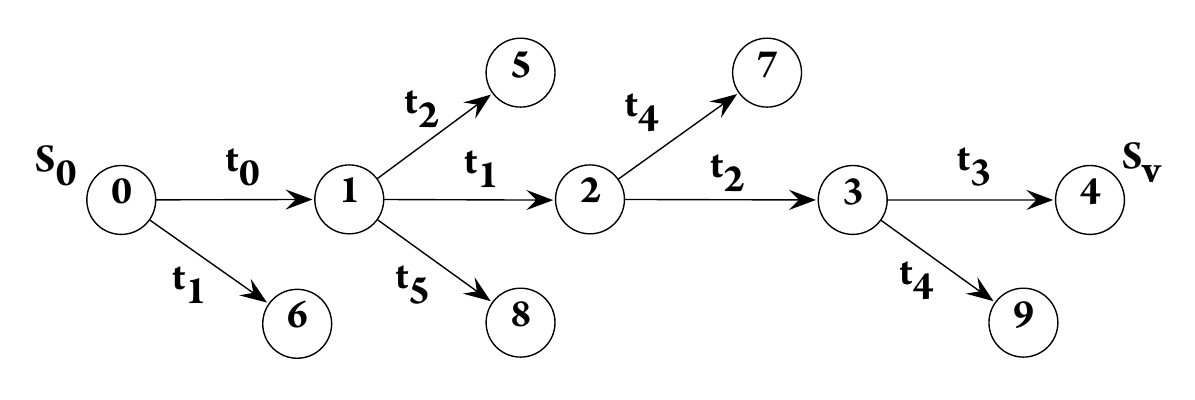}
    \caption{Error Trace Example}
    		\label{7fig:errortraceex}
\end{figure}
\end{example}

\subsection{Kullback-Lerbler Divergence}
\dkl (also called information divergence, information gain, relative entropy) \cite{kullback1951information} is a fundamental equation of information theory that qualifies the proximity of two probability distributions. 

\begin{definition}[\dkl (\texttt{KL})]
KL Divergence is a measure in statistics that quantifies in bits how close a probability distribution \\ $P = \{p_i\}$ is to a model (or candidate) distribution Q = $\{q_i\}$. The KL-divergence of Q from P over a discrete random variable is defined as
\begin{equation}
D_{KL} (P \parallel Q) = \sum\limits_{i}  P(i) \ln \frac{P(i)}{Q(i)}
\end{equation}
Note: In the definition above, $0 \ln \frac{0}{0} = 0$,  $0 \ln \frac{0}{q} = 0$, and $p \ln \frac{p}{0} = \infty$. 
\end{definition}

\dkl has many applications. We give an example of its application to text classification \cite{baker1998distributional}.  A textual document $d$ is a discrete distribution of $|d|$ random variables, where $|d|$ is the number of terms in the document. Let $d_1$ and $d_2$ be two documents whose similarity we want to compute. This is done using $D_{KL} (d_1 \parallel d_2)$ and $D_{KL} (d_2 \parallel d_1)$.

Another major application is the \tf-\idf (Term Frequency - Inverse Document Frequency) algorithm \cite{jones1972statistical}. \tf-\idf is a numerical statistic which reflects how important a term is for a given document in a corpus (collection) of documents. It is often used as a weighting factor in information retrieval and text mining.  Variations of the \tf-\idf weighting scheme are often used by search engines as a central tool in scoring and ranking a document's relevance to a given user query.

Suppose we have a collection of English textual documents and aim to determine which documents are most relevant to the query "the model checking". We might start by eliminating documents that do not contain the three words "the", "model", and "checking", but this still leaves many documents. To further distinguish them, we might count the number of times each term occurs in each document and sum them all together; the number of times a term occurs in a document is called its term frequency (\tf).

However, because the term "the" is so common, this might incorrectly emphasize documents which happen to use the word "the" more frequently, without giving enough weight to the more meaningful terms "model" and "checking". The term "the" is not a good keyword to distinguish relevant and non-relevant documents and terms, unlike the less common words "model" and "checking". Hence an inverse document frequency (\idf) factor is incorporated which diminishes the weight of terms that occur very frequently in the document set and increases the weight of terms that occur rarely.


\subsection{Ranking Suspicious Faulty Transitions}
Inspired by the \tf-\idf algorithm, we propose a probabilistic fault localization approach based on the \dkl. A relevance weight $C_F(t)$ is computed to assess the contribution of a transition $t$ in error traces leading to violation states and thus its contribution to the fault.

\begin{figure}[ht!]
	\centering
		\includegraphics[height=0.1\textheight]{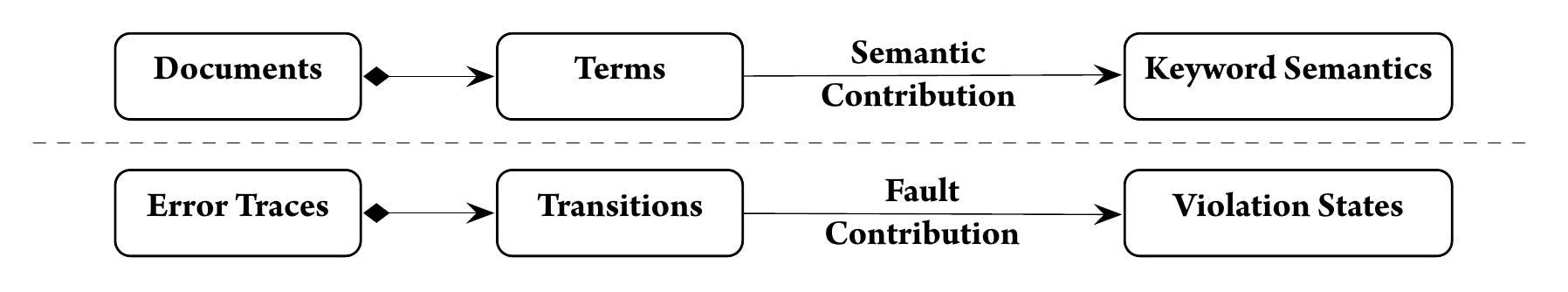}
    \caption{Comparison to \tf-\idf}
    		\label{7fig:coreidea}
\end{figure}

In the \tf-\idf algorithm, each term in the documents will contribute to the semantics of keywords. Some terms are considered as significant if they are more relevant to the semantics of keywords. This is similar to the fault contribution caused by a given transition in an error trace in model checking. Fig. \ref{7fig:coreidea} compares the similarity between semantic contribution of terms in documents and fault contribution of transitions in error traces. Some terms in documents have closer semantic relation to the keywords, the occurrence of these terms provide more semantic contributions to the occurrence of keywords. Similarly, the fault propagation depends on the topology of error traces, the occurrence of some transitions will provide more fault contributions to the occurrence of violation states.

The semantic contribution of a term in documents is measured by \tf-\idf, where \tf is the contribution of a term in single document, and \idf is the contribution of a term in a collection of documents. The fault contribution to the violation states $\{s_{vi}\}$ caused by a transition $t$ on error traces $\{\pi_i\}$ can also be evaluated by a similar measure $C_F(t)$, defined as \tc-\itc (Transition Contribution - Inverse Trace Contribution). $C_F(t) = \tc(t) \cdot \itc(t)$.

\subsection{Experimental Results}
We assess our approach using two significant criteria: effectiveness and efficiency.
According to the survey \cite{wong2009survey}, the effectiveness can be assessed by a score \exam in terms of the percentage of statements that have to be examined until the first statement containing the fault is reached \cite{eric2010family,wong2009bp}. 
The fault localization techniques in model checking, like other techniques, should terminate in a timely manner, limited by some resource constraints. The efficiency can be assessed by the scalability and the performance.


\subsubsection{Automated Test Bed}
The test bed will generate randomly systems which might have deadlock, then apply the proposed analysis algorithm and check that it detects the introduced deadlocks. We use Time Petri Net to model system's behavior. 

For a given \tpn system $S(P, R, M)$, $P$ are the processes which run infinitely and need a resource before the next task (a task is represented by a transition); $R$ are resource which are shared by all the processes, but only accessible in an exclusive way; $M$ is a matrix to decide whether process $P_i$ will need to access resource $R_j$.  Coffman identified four conditions that must hold simultaneously in order to have a deadlock \cite{coffman1971system}. 
To improve the success of creating a deadlock in the system, we introduced another mechanism to enforce deadlocks: randomly let some processes during some tasks forget to release a used resource. These tasks are then considered as the error source of system's deadlock. 
\subsubsection{Evaluation of Efficiency}
We have generated thousands of test cases by assigning $P$ and $R$ values from 5 to 20, creating 1 to 9 faulty transitions, with all the other parameters totally random. 
The tests are performed on a 2,4 GHz Intel Core 2 Duo processor running Mac OS X 10.6.8.  The system parameters and efficiency evaluation results are shown in Table \ref{7tab:evaluationefficiency}. The average time of evaluation shows that the approach is efficient for large scale system.

\begin{table}[ht!]
\scriptsize
	\caption{Efficiency Evaluation}
    \begin{center}
    \begin{tabular}{| c | c | c | c |}
  	\hline
  	\multicolumn{3}{|c|}{\textbf{System}} & \textbf{Evaluation}  \\
  	\hline
  	\textbf{Fault Num.} & \textbf{Test Num.} & \textbf{Av. State/Transition} & \textbf{Average Time (s)} \\ 
	\hline
	\hline
	1 & 400 & 4949 / 15440  & 2.9092 \\
	\hline
	2 & 517 & 2428 / 7130  & 1.1244 \\
	\hline
	3 & 500 & 9884 / 31237 &  3.3533 \\
	\hline	
	4 & 402 & 8811 / 26663 &  2.5998   \\
	\hline	
	5 & 303 & 6756 / 18247 &   1.2196 \\
	\hline	
	6 & 504 & 27094 / 75808 &  5.064 \\
	\hline	
	7 & 757 & 104857 / 304741 &  15.0072  \\
	\hline	
	8 & 100 &  112306 / 283004 &   15.0289 \\ 
	\hline
	9 & 1 & 241920 / 583200 &   36.906 \\
	\hline
	\end{tabular}
    \end{center}
\label{7tab:evaluationefficiency}
\end{table}

\subsubsection{Evaluation of Effectiveness}
The effectiveness evaluation is shown in Table \ref{7tab:evaluationeffectiveness}. We give out \exam score, \exam score variance, rank, and rank variance for the best cases and worst cases, and then show the average \exam score and average rank. The \exam score varies from 2\% to 13\% for best cases, and varies from 4\% to 18\% for worst cases. In average, \exam varies from 3\% to 16\% which corresponds to ranking results from 1 to 8. The stability is represented by the variance result. These experimental results shows our approach is effective.

\begin{table}[ht!]
\scriptsize
	\caption{Effectiveness Evaluation}
    \begin{center}
    \begin{tabular}{| c | c | c | c | c | c | c | c | c | c | c |}
  	\hline
  	\multirow{2}{*}{\textbf{F. N.}} & \multicolumn{4}{|c|}{\textbf{Best Case}} & \multicolumn{4}{|c|}{\textbf{Worst Case}} & \multicolumn{2}{|c|}{\textbf{Average}} \\
  	\cline{2-11}
  		& \textbf{\exam} & \textbf{\exam Var} & \textbf{Rank} & \textbf{Rank Var} & \textbf{\exam} & \textbf{\exam Var} & \textbf{Rank} & \textbf{Rank Var} & \textbf{\exam}  & \textbf{Rank} \\ 
	\hline
	\hline
	1	&	0,13335	&	0,00134	&	3,25	&	1,79	&	0,18603	&	0,00244	&	4,33	&	1,63	&	0,15969	&	3,79	\\
	\hline
	2	&	0,04229	&	0,00219	&	1,1	&	1,75	&	0,09574	&	0,00213	&	2,11	&	1,75	&	0,069015	&	1,605	\\
	\hline
	3	&	0,02108	&	0,00106	&	0,75	&	1,52	&	0,05892	&	0,0009	&	1,75	&	1,52	&	0,04	&	1,25	\\
	\hline
	4	&	0,00722	&	0,0004	&	0,26	&	0,49	&	0,039	&	0,00042	&	1,26	&	0,49	&	0,02311	&	0,76	\\
	\hline
	5	&	0,02044	&	0,0017	&	0,83	&	2,95	&	0,0478	&	0,00162	&	1,83	&	2,95	&	0,03412	&	1,33	\\
	\hline
	6	&	0,05369	&	0,00336	&	2,46	&	7,36	&	0,0766	&	0,0033	&	3,46	&	7,36	&	0,065145	&	2,96	\\
	\hline
	7	&	0,08857	&	0,00372	&	4,61	&	10,9	&	0,10822	&	0,0037	&	5,61	&	10,9	&	0,098395	&	5,11	\\
	\hline
	8	&	0,13091	&	0,00099	&	7,3	&	3,95	&	0,14905	&	0,001	&	8,3	&	3,95	&	0,13998	&	7,8	\\
	\hline
	9	&	0,10169	&	0	&	6	&	0	&	0,11864	&	0	&	7	&	0	&	0,110165	&	6,5	\\
	\hline
	\end{tabular}
    \end{center}
\label{7tab:evaluationeffectiveness}
\end{table}

\section{Probabilistic Failure Analysis in Simulation}
\label{simulation}
\subsection{Background}
Fault localization algorithms usually follow two paradigms: cause-effect and effect-cause analyses. Cause-effect analysis \cite{takahashi2002diagnosing,zeller2002isolating,jobstmann2012finding} starts from possible causes (fault models). A simulator is used to predict system's behavior in the presence of various faults. Then predictions are matched against observed behavior. Effect-cause analysis \cite{abramovici1980multiple,smith2005fault} reasons faulty localization based on observed behavior and expected good functions. It back-traces faulty causes from the identified suspect components. 

In this work, we make a trade-off of cause-effect and effect-cause analyses and propose an Hidden Markov Model (\hmm) \cite{rabiner1989tutorial} based approach for the automated localization of faulty components in the simulation \cite{ge2013hidden,ge2013efficient,ge2015online}. The component can be hardware device, software modules or functional blocks in the system. 
This method combines forward localization analysis and backward confidence degree evaluation. \hmm, as a component's abstraction, provides statistically identical information to component's real behavior. The core of this method is a fault localization algorithm that gives out the set of suspicious ranked faulty components and a backward algorithm that computes the matching degree between the \hmm and the simulation model to evaluate the confidence degree of the localization conclusion. 

\subsection{HMM Modeling and Analysis}
\label{sec:hmm}
An \hmm is defined as a statistical model used to represent stochastic processes, where the states are not directly observed. 
A basic \hmm can be described as follows:
\begin{itemize}
\item
$\mathbf{N}$: number of states
\item
$\mathbf{M}$: number of observations
\item
$\mathbf{M_I}$: initial probability distribution; $\displaystyle\sum\limits_{i=1}^{N} \mathbf{M_I}(i) = 1$
\item
$\mathbf{M_T}$: probability distribution of transitions from states to states; \\ $\displaystyle\sum\limits_{j=1}^{N} \mathbf{M_T}(i,j) = 1, i = 1...N$
\item
$\mathbf{M_E}$: emission distribution for the observations associated with states;  \\ $\displaystyle\sum\limits_{j=1}^{M} \mathbf{M_E}(i,j) = 1, i = 1...N$
\end{itemize}

\begin{example}[HMM Example]
\label{ex:hmm}
A two states \hmm example abstracting a system's health condition is given by Fig. \ref{fig:hmm}, where the system owns two states \textit{Healthy ($\mathcal{H}$)} and \textit{Faulty ($\mathcal{F}$)}, and two observations which represent the ouputs respect the functional constraints ($\mathcal{R}$) or violate the functional constraints ($\mathcal{V}$). The three distributions $\mathbf{M_I} \mathbf{M_T}  \mathbf{M_E}$ are:
$\begin{pmatrix}
		0.6 & 0.4
\end{pmatrix}
\begin{pmatrix}
		0.7 & 0.3 \\
		0.4 & 0.6
\end{pmatrix}
\begin{pmatrix}
		0.9 & 0.1 \\
		0.2 & 0.8
\end{pmatrix}$.
\begin{figure}[h!]
	\begin{center}
	\includegraphics[width=0.38\textwidth]{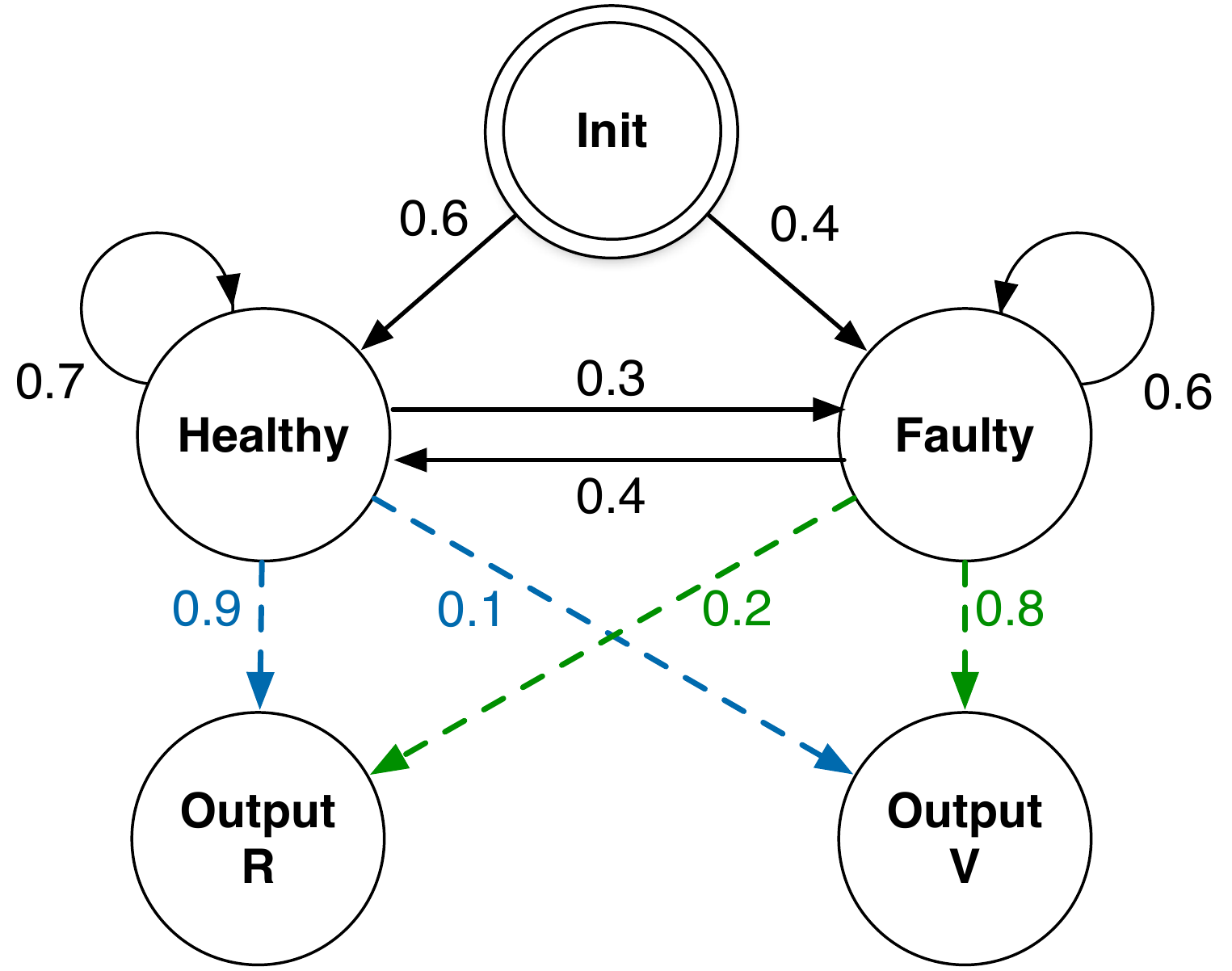}
	\caption{A Two States Hidden Markov Model}
	\label{fig:hmm}
	\end{center}
\end{figure}
\end{example}

\hmm, as abstract model of real system, is statistically identical to system's real behavior. 
When modeling a system, \hmm separates the concept into two conceptually independent paradigms: behavior and observation. Behavior refers to what the system really is; while observation to what the system exhibits that is used for its recognition. $\mathbf{M_I}$ gives indication about the probability that a behavior becomes the first behavior when system runs. $\mathbf{M_T}$ decides how probably will the system behave from one state to the other states. This is statistically equivalent to the real system's behavior. $\mathbf{M_E}$ provides a distribution that  connects the behavior and the observation: if at a given time the behavior is known, how probably an observed sequence will occur. 

$\mathbf{M_I}, \mathbf{M_T}$ and $\mathbf{M_E}$ can be obtained by modeling or through a learning process. Once all these matrix parameters are estimated, \hmm is capable to deduce, given an observed output sequence or a set of such sequences, the maximum likelihood estimation of inner-state transition sequences.

\subsection{Automated Fault Localization Based on Hidden Markov Model}
\begin{figure}[h!]
	\centering
		\includegraphics[height=0.2\textheight]{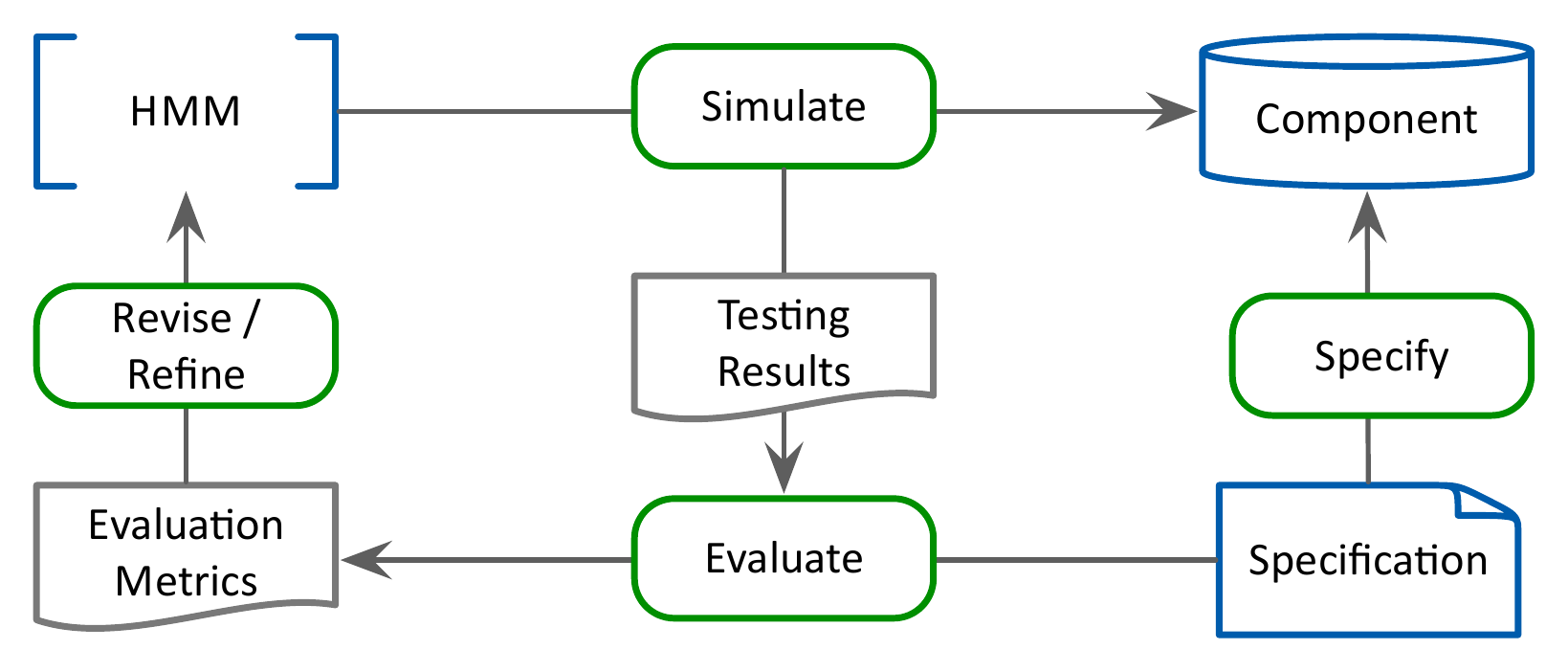}
    \caption{HMM Fault Localization Framework}
    		\label{fig:localframework}
\end{figure}

This approach is based on component analysis. Each component is mapped to an \hmm. If all the input/output pair of a component can be exhaustively listed, we can get an exact distribution of how this component respects the functional constraints. This approach can be explained by using Fig. \ref{fig:localframework}. An \hmm, as a component's abstraction, provides statistically similar information to simulation by $\mathbf{M_I}$. That is equally saying, a component can be simulated by \hmm if we can be sure they behave statistically in the same way. To measure whether they behave the same, we introduce an evaluation approach by using test results. The returned evaluation metrics are used to revise or refine the parameters in \hmm, until it approximates the component's real behavior.

\subsubsection{System States}

A component C is mapped to an \hmm. Hmm's states are the combination of component's faulty status and its inputs faulty status. System's behavior is modeled by 4 states:

\begin{itemize}
\item
$PI_p$: C is not faulty, C's inputs are passed
\item
$PI_f$: C is not faulty, C's inputs are failed
\item
$FI_p$: C is faulty, C's inputs are passed
\item
$FI_f$: C is faulty, C's inputs are failed
\end{itemize}

\subsubsection{Observation $\&$ Observed Sequence}
The observation is defined by the test results of component's outputs. There are only 2 observations:
\begin{itemize}
\item
$O_p$: outputs are passed
\item
$O_f$: outputs are failed
\end{itemize}

The dependency between components is built By defining \hmm states and observations in this way. For component C, its output observation is the input (contained in the defined states) of the successor.

\subsubsection{Initial Probability Matrix $\mathbf{M_I}$}
Assume components' faulty probability is $\omega$, $\mathbf{M_I}$ is defined as following table \ref{tab:initmatrix}. If $\omega$ is not available, we assume each state has identical initial faulty probability, i.e. $\omega = \frac{1}{2}$. 

\begin{table}[ht]
\begin{minipage}[b]{0.45\linewidth}\centering
	\caption{Initial Probability Matrix $M_I$}
\begin{tabular}{| c | c c c c |} 
    \hline
		&  $PI_p$ & $PI_f$ & $FI_p$ & $FI_f$ \\
      \hline
     Init & $\frac{1 - \omega}{2}$ & $\frac{1 - \omega}{2}$ & $\frac{\omega}{2}$ & $\frac{\omega}{2}$ \\
      \hline
    \end{tabular}
    \label{tab:initmatrix}
\end{minipage}
\hspace{0.5cm}
\begin{minipage}[b]{0.45\linewidth}
\centering
    \caption{Transition Probability Matrix $M_T$}
\begin{tabular}{| c | c c c c |} 
    \hline
		&  $PI_p$ & $PI_f$ & $FI_p$ & $FI_f$ \\
      \hline
     $PI_p$ & $\frac{\alpha (1 - \omega)}{\alpha + \beta}$ & $\frac{\beta (1 - \omega)}{\alpha + \beta}$ & $\frac{\alpha \omega}{\alpha + \beta}$ & $\frac{\beta \omega}{\alpha + \beta}$ \\
     $PI_f$  & $\frac{\gamma (1 - \omega)}{\gamma + \delta}$ & $\frac{\delta (1 - \omega)}{\gamma + \delta}$ &  $\frac{\gamma \omega}{\gamma + \delta}$ & $\frac{\delta \omega}{\gamma + \delta}$ \\
     $FI_p$  & $\frac{\alpha (1 - \omega)}{\alpha + \beta}$ & $\frac{\beta (1 - \omega)}{\alpha + \beta}$ & $\frac{\alpha \omega}{\alpha + \beta}$ & $\frac{\beta \omega}{\alpha + \beta}$ \\
     $FI_f$ & $\frac{\gamma (1 - \omega)}{\gamma + \delta}$ & $\frac{\delta (1 - \omega)}{\gamma + \delta}$ & $\frac{\gamma \omega}{\gamma + \delta}$ & $\frac{\delta \omega}{\gamma + \delta}$ \\
      \hline
    \end{tabular}
    \label{tab:transitionmatrix}
\end{minipage}
\end{table}

\subsubsection{Transition Probability Matrix $\mathbf{M_T}$}
$\mathbf{M_T}$ contains statistical values derived from test results. If $n$ test cases are observed, a component with $m$ inputs corresponds to $m \cdot n$ input sequences. Using the $m \cdot n$ input sequences, we can compute the fault probabilities $\alpha, \beta, \gamma, \delta$ respectively for the transitions the states, where $\alpha + \beta + \gamma + \delta = 1$. $\mathbf{M_T}$ is then calculated as following table \ref{tab:transitionmatrix}. If $\omega$ is not available, we assume $\omega = 1$.

\subsubsection{Emission Probability Matrix $\mathbf{M_E}$}
The key of this approach is evaluating $M_E$. According to previous analysis, we need to measure how well an \hmm statistically simulates a given component, which is represented by \textit{Matching Level}.

\begin{definition}[Matching Level ($\mu$)]
Matching level evaluates how well \hmm simulates a component's real behavior.
\end{definition}

The objective is to find an \hmm with the highest matching level. This turns the problem to be an optimization problem, and many techniques can be applied, e.g. exhaustive search by minimum internal, heuristic algorithm, evolution algorithm, experimental design, etc. In this work, the matching level is derived by calculating the probability that a component's outputs pass the tests through \hmm observations. However, this matching level is a local value relative to one component, making it have no meaning to compare with others. Therefore, the forward search only is not enough to guarantee the matching level. We introduce the concept of \textit{Confidence Level} to evaluate the global confidence of matching level.

\begin{definition}[Confidence Level ($\rho$)]
Confidence level evaluates the confidence of the matching level.
\end{definition}

We propose the following algorithm to evaluate the $M_E$ in \hmm (h) corresponding to component C. $T_{m,n} = \{\tau_1, \tau_2, ..., \tau_n \}_m$ are test cases results for m outputs of C. The threshold of matching level $\mu$ and confidence level $\rho$ are pre-defined. 

\begin{figure}[ht!]
	\centering
		\includegraphics[height=0.3\textheight]{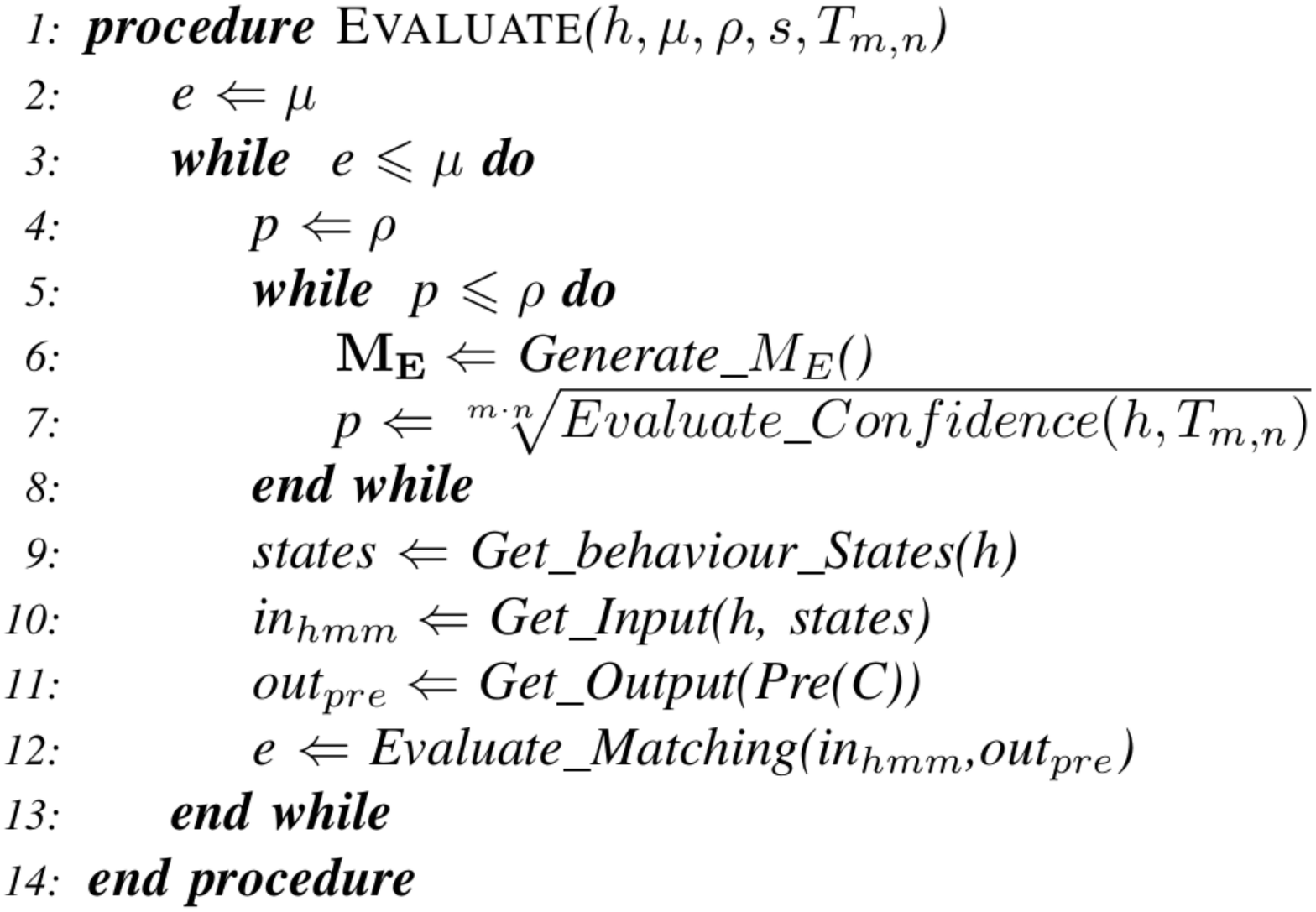}
    		\label{fig:me}
\caption{Algorithm: evaluation of $M_E$}
\end{figure}

\subsection{Estimating System's Behavior $\&$ Locating Fault}
When the \hmm with high matching level and high confidence level is confirmed, we can estimate the component's status. Using the states calculated from the function $\mathtt{Get\_Behaviour\_States(h, T_{m,n})}$, e.g. a states behavior sequence is derived, an example of which is given   
$\mathtt{(FI_p, FI_p, PI_f, FI_p, ... FI_f, FI_p, ...)}$.

In the states behavior sequence, we focus on the status of component, shown as follow: $\mathtt{(F, F, P, F ..., F, F, ...)}$. The status with higher occurrence probability is confirmed as this component's faulty status.

This result's confidence is guaranteed by the confidence level ($\rho$), and the \hmm's similarity to the component is guaranteed by the matching level ($\mu$). $\rho$ and $\mu$ are computed within system's topological structure, therefore we can compare all the faulty component's $\rho$ and $\mu$, and give a set of suspect faulty components ordered by $\rho$ and $\mu$.

\subsection{Experimental Results}
We design a specific test bed to assess the method's accuracy and efficiency by generating a large number of use cases. Each use case includes: 
the \textbf{system architecture} which defines the components and the ports, and their interconnections;
the \textbf{failure probability} of each component;
the \textbf{functional specification} corresponding to the inputs/outputs.

The method assumes that each component in the system has a chance to fail if it has design problems. This probability will be 0 if no design fault is presumed for this component. All functional constraints are based on input/output's value itself, and for simplification, they are all range constraint, which means they delimit only the min/max value of the input/output.
If a faulty component exists, the test bed will by chance give out an out-of-range value for this component's output. This emulates how a system fails, whatever the model is.
Each component's input and output will be allocated to a variable by the test bed. It guarantees that the interconnected ones share the same variable. The variable will be associated with a random range, which is the functional specification. If a device is more probable to fail, the test-bed-generated value for its entire output variables will be more probable to go out of the defined range.
The approach will use the generated data and the functional specification to automatically locate the faulty component. The test bed will then compare this computed conclusion with the initial context to deduce whether the method is efficient.

\begin{figure}[ht!]
	\centering
		\includegraphics[height=0.25\textheight]{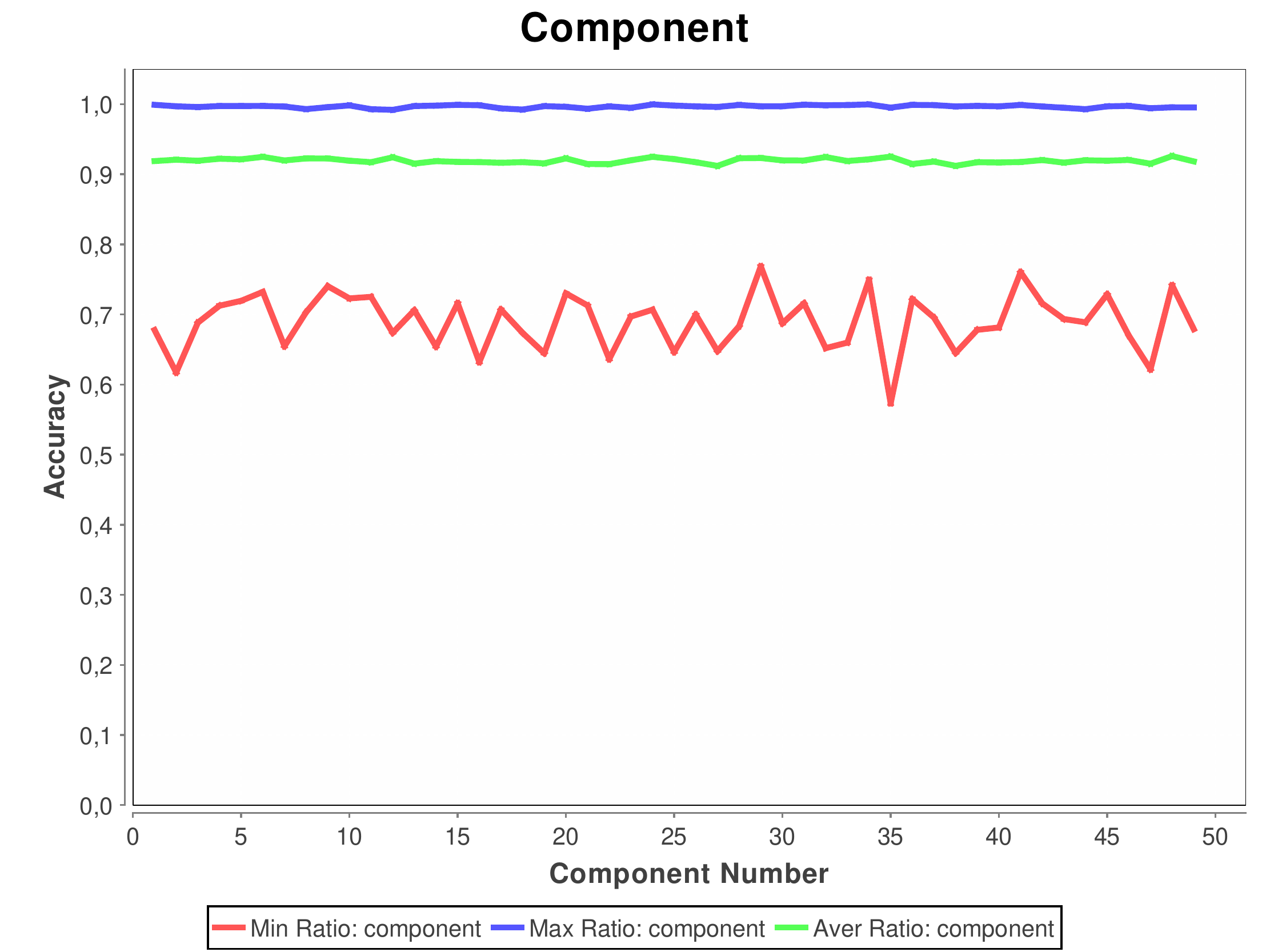}
    \caption{Accuracy by Component Number}
    		\label{fig:comp}
\end{figure}

\begin{figure}[ht!]
	\centering
		\includegraphics[height=0.25\textheight]{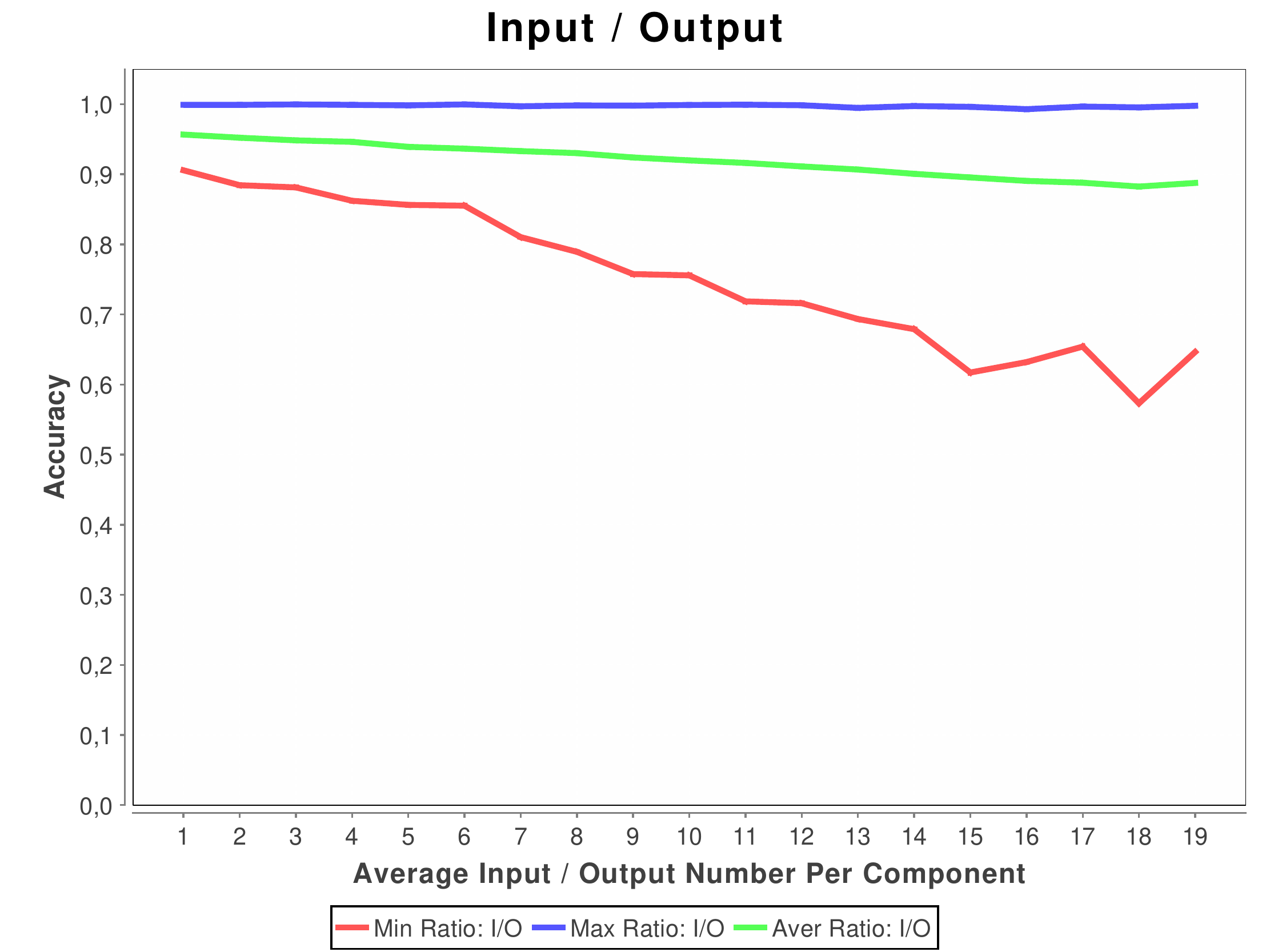}
    \caption{Accuracy by Input number}
    		\label{fig:input}
\end{figure}

The test bed generated 1000 use cases to assess the performance in terms of accuracy. The criteria that impacts the accuracy is the complexity of system's architecture. This can be measured by component number (Fig. \ref{fig:comp}). and component's average input \& output number (Fig. \ref{fig:input}). We find out that this method is more sensible to the average input \& output number, while it is more scalable to component number. This method deals with the fault localization for middle-range systems with a accuracy superior to 90\%.

\vspace*{-2cm}

The computation of ME parameters by iteratively searching algorithm consumes time, which is bounded by the iteration limit. This Monte-Carlo algorithm runs from several seconds to several minutes. However, as the computation of each component is independent, the whole method is linearly scalable. For a large system, a parallel cluster will locate the probable design faults within minutes.

\section{Conclusion}
\label{conclusion}
Automated fault localization is an important issue in model \vv. It helps the end users in analyzing the origin of failure. 
In this work, we have shown the early experiments with probabilistic analysis approaches in fault localization. Inspired by the  \dkl from Bayesian probabilistic theory, we propose a suspiciousness factor to compute the fault contribution for the transitions in the reachability graph of model checking. The potential faulty transitions are then ranked according to this suspiciousness factor. To automatically locate design faults in the simulation model of detailed design, we propose to use the statistical model Hidden Markov Model (\hmm).  \hmm, as a component's abstraction, provides statistically identical information to component's real behavior. The core of this method is a fault localization algorithm that gives out the set of suspicious ranked faulty components and a backward algorithm that computes the matching degree between the \hmm and the simulation model to evaluate the confidence degree of the localization conclusion. 

\section*{Acknowledgment}
This work was funded by the FUI Projet P and EuroStars HiMoCo projects. 
\bibliographystyle{splncs03}
\bibliography{report}
\end{document}